\documentclass[12pt]{article}
\usepackage{amsfonts}
\begin{document}
\begin{flushright}
LPT/ENS-03/38, \\ FTUV-03-1228 
\\ ESI-1430 \\ hep-th/0312299,
\\ December 28, 2003 
\end{flushright}

\begin{center}
{\Large Superfield T-duality  rules in ten dimensions with one isometry
\footnote{{\small\it To be published in the Proceedings
 of the International Workshop "Supersymmetries and Quantum
 Symmetries" (SQS'03, 24-29 July, 2003), JINR Publishing, Dubna, Russia.}}}

\medskip 

{\bf Igor A.~Bandos$^{\dagger\ast}$ and Bernard Julia$^{\ddagger}$}

\medskip

{\small\it $^{\dagger}$
Departamento de F\'{\i}sica Te\'orica and IFIC (CSIC-UVEG)\\
    46100-Burjassot (Valencia), Spain, \\
 $^{\ddagger}$
        Laboratoire de Physique Th\'eorique de l'Ecole Normale Sup\'erieure\\
    75231 Paris Cedex 05, France, \\
$^{\ast}$
    Institute for Theoretical Physics, NSC KIPT
    UA61108, Kharkov, Ukraine }

\end{center}

\begin{abstract}
In this contribution we present the superfield T-duality
rules relating type IIA and type IIB supergravity potentials for
the case when both type IIA and type IIB superspaces have (at
least) one isometry direction. We also give a brief review of
T--duality and discuss the main steps of our approach to the
derivation of the superfield T--duality rules, including the
treatment of T-duality as an operation acting on differential
forms rather than on the superspace coordinates.
\end{abstract}

\section{Introduction}

T--duality is a perturbative symmetry of closed string theory
which appears when at least one of the spacetime directions is
compact (see, {\it e.g.}, \cite{Polch0}).

The string equations of motion may be linearized by fixing a
conformal gauge for the worldsheet metric, $\Box {} \tilde{X}^\mu
(\tau , \sigma) = 0$. Then the general solution describes
independent motions of right--movers ($\tilde{X}_R^\mu$)
and left--movers ($\tilde{X}_L^\mu$)
\begin{eqnarray}\label{StrEq}
\Box {} \tilde{X}^\mu (\tau , \sigma) = 0 \quad \Rightarrow
\quad  \tilde{X}^\mu (\tau , \sigma) = \tilde{X}_L^\mu (\tau -\sigma)
+ \tilde{X}_R^\mu (\tau + \sigma) \; ; \\ \nonumber
left\; movers \; : \quad
\tilde{X}_L^\mu (\tau - \sigma) = (\tau -\sigma) p_L^\mu  + \,
oscillating \;  terms\; , \quad \\ \nonumber right\; movers: \quad
\tilde{X}_R^\mu (\tau + \sigma) = (\tau +\sigma) p_R^\mu  + \,
oscillating  \; terms\; , \quad
\end{eqnarray}
In flat noncompact spacetime the periodic boundary conditions in
$\sigma$ ($\tilde{X}^\mu (\tau , \sigma)= \tilde{X}^\mu (\tau ,
\sigma + 2\pi )$) implies $ p_L^\mu=  p_R^\mu= 1/2 p^\mu$. If some
direction, say $X^{25}$, is compact, $X^{25} \sim X^{25} + 2\pi m
$, $m \in \mathbf{Z}$, then, firstly, the momentum $p^{25}=
p^{25}_R +p^{25}_L$ becomes discrete $ p^{25} \in \mathbf{Z}$,
and, secondly, the difference $p_R^\mu -  p_L^\mu$ is integer
rather than vanishing as in the noncompact case. This integer,
$p_R^\mu - p_L^\mu = n^{25} \in \mathbf{Z}$ is called the `winding
number'. It counts the number of times the string wraps the
compact $X^{25}$ direction. Thus, in this case,
\begin{eqnarray}\label{X25s}
\tilde{X}^{25} (\tau , \sigma) = p^{25} \tau +  n^{25} \sigma \, +
\, oscillating  \; terms\; , \quad p^{25},  n^{25} \in
\mathbf{Z}\; ,
\end{eqnarray}
and the T-duality is an operation which interchanges the momentum number
$ p^{25} \in \mathbf{Z}$ and the winding number $ n^{25} \in
\mathbf{Z}$,
\begin{eqnarray}\label{pTw}
 p^{25} & \longleftarrow{}^{T-duality}\longrightarrow & n^{25} \; .
\end{eqnarray}
More precisely T-duality changes the sign of $\partial
\tilde{X}^{25}_L$.

In open string theory the T-duality,  as defined above, maps
the Neumann boundary conditions in the isometry direction
(corresponding to the string with
free ends) into the Dirichlet boundary conditions (corresponding
to the string endpoints attached to some hyperplane),
\begin{eqnarray}\label{NTD}
\begin{matrix} {Neumann \cr boundary \; condition}\end{matrix} &
\longleftarrow{}^{T-duality}\longrightarrow & \begin{matrix}
{Dirichlet \cr boundary \, condition  }\end{matrix}\; . \quad
\end{eqnarray}
As the string theory implies gravity, the hyperplane where a
string can have its endpoints cannot be treated as "frozen", but
it is rather a dynamical object;  this
 is identified as a so--called Dirichlet $p$--brane or D$p$--brane
\cite{Polch}). Thus, in the case of spacetime with (at least)
one isometry direction, T-duality should map a D$p$--brane
onto either a D$(p-1)$--brane or a D$(p+1)$--brane, depending on whether the
isometry direction is tangential or orthogonal to the
D$p$--brane worldvolume $W^{p+1}$,
\begin{eqnarray}\label{pTppm1}
 Dp-brane  & \longleftarrow{}^{T-duality}\longrightarrow & D(p\pm
1)-brane \; .
\end{eqnarray}

T-duality is compatible with spacetime supersymmetry.
In flat superspace T-duality can be defined as a map between
type IIA and type IIB superstring models
that transforms nontrivially one of the fermionic coordinate
functions
\begin{eqnarray}\label{IIATIIB}
\begin{matrix}{type\; IIA \cr
superstring }\end{matrix}
 \; \left(\begin{matrix}{
\tilde{\hat{\theta}}{}^{\alpha 1} (\tau , \sigma)\cr
\tilde{\hat{\theta}}_{\alpha}{}^2  (\tau , \sigma)}
\end{matrix} \right)& \longleftarrow{}^{T-duality}\longrightarrow
& \left(\begin{matrix}{\tilde{{\theta}}{}^{\alpha 1}  (\tau ,
\sigma)\cr \sigma^{\#}_{\alpha\beta} \tilde{{\theta}}{}^{\beta 2}
(\tau , \sigma) }\end{matrix}\right) \; \begin{matrix}{type \; IIB
\cr superstring }\end{matrix}
\end{eqnarray}
where the isometry direction is identified with $\hat{z}=
\hat{X}^{9}$ and $y=X^{9}$, respectively, and $
\sigma^{\#}_{\alpha\beta}= \sigma^{9}_{\alpha\beta}$ is the $D=10$
Majorana--Weyl gamma--matrix related to the isometry direction.
Eq. (\ref{IIATIIB}) suggests that the T-duality should transform
the fermionic coordinates of {\it flat} type
 IIA and type IIB superspaces with topology of the bosonic sector
$\mathbf{R}^{9}\otimes S^1$ (see, {\it e.g.} \cite{simon2} and
refs. therein)
\begin{eqnarray}\label{thIIATIIB}
\left(\begin{matrix}{ {\hat{\theta}}{}^{\alpha 1} \cr
{\hat{\theta}}_{\alpha}{}^2 }
\end{matrix} \right)& \longleftarrow{}^{T-duality}\longrightarrow
& \left(\begin{matrix}{{\theta}{}^{\alpha 1} \cr
\sigma^{\#}_{\alpha\beta} {{\theta}}{}^{\beta 2}
}\end{matrix}\right) \; .
\end{eqnarray}

\section{
T-duality  rules for bosonic fields}

In 1987  Buscher found that the study of the string action in a
bosonic background with isometries allows one to find an elegant
field theoretical representation for T-duality  as a relation
between {\it bosonic NS--NS} ({\it
Neveu-Schwarz--Neveu-Schwarz}) fields of type IIA and type IIB
supergravity theories:
\begin{eqnarray}\label{NS-NS}
\nonumber NS-NS\;\;\; type \; IIA \quad & & NS-NS\;\;\; type\; IIB
\\
 \hat{g}_{\hat{\mu}\hat{\nu}}(\hat{x})\, ,
 \hat{B}_{\hat{\mu}\hat{\nu}}(\hat{x})\, , \, \, \hat{\phi}(\hat{x})\, , \,
& \leftarrow^{T-duality} \rightarrow &  {g}_{{\mu}{\nu}}({x})\, ,
 \hat{B}_{{\mu}{\nu}}({x})\, , \, \, {\phi}({x})\, . \,
\end{eqnarray}
For the case of spacetime(s) with one isometry they have the form
\footnote{We use the metric in the so--called 'Einstein
frame', where the Einstein--Hilbert  action does not include a
dilaton factor, while the string action does. This results in the
occurence of the dilaton factors in Eqs.
(\ref{TrA})--(\ref{TdilatB}).}
\begin{eqnarray}\label{TrA}
& e^{{\phi\over 2}}g_{yy}={1\over e^{{\hat{\phi}\over
2}}\hat{g}_{\hat{z}\hat{z}}}\; , \qquad e^{{\phi\over 2}}
{g}_{\tilde{ {m}}y}= {1\over e^{{\hat{\phi}\over 2}}
\hat{g}_{\hat{z}\hat{z}}} \hat{B}_{\hat{z}\tilde{ {m}}}\; , &
\qquad    \\
\label{TrulesA-} & e^{{\phi\over 2}} g_{\tilde{ {m}}\tilde{ {n}}}
= e^{{\hat{\phi}\over 2}} \hat{g}_{\tilde{ {m}}\tilde{ {n}}} +
{1\over e^{{\hat{\phi}\over 2}} \hat{g}_{\hat{z}\hat{z}}} \left (
{\hat{B}}_{\tilde{ {m}} \hat{z}} {\hat{B}}_{\tilde{ {n}} \hat{z}}
- e^{\hat{\phi}}\hat{g}_{\tilde{ {m}} \hat{z}} \hat{g}_{\tilde{
{n}} \hat{z}}\right)\; , & \qquad
\\ \label{TdilatB}
& e^{2\phi} =- {e^{2\hat{\phi}}\over e^{{\hat{\phi}\over 2}}
\hat{g}_{\hat{z}\hat{z}}}\; ,  & \qquad
\\
\label{T-Bb} & B_{\tilde{{ {m}}}\tilde{{ {n}}}} =
\hat{B}_{\tilde{{ {m}}}\tilde{{ {n}}}} + {1\over
\hat{g}_{\hat{z}\hat{z}}} \left ( \hat{g}_{\tilde{{ {m}}}\hat{z}}
\hat{B}_{\tilde{ {n}} \hat{z}} - \hat{g}_{\tilde{ {n}} \hat{z}}
\hat{B}_{\tilde{ {m}} \hat{z}} \right),  \quad  & B_{y\tilde{
{m}}} =
 {1\over \hat{g}_{\hat{z}\hat{z}}}
\hat{g}_{\tilde{ {m}} \hat{z}}\; .  \qquad
\end{eqnarray}
Here $\partial_{\hat{z}}$ is along the isometry direction of the
type IIA supergravity which is defined on the space with the
coordinates $\hat{x}^{\hat{\mu}}= (x^{\tilde{m}}, \hat{z})$ (which
may be called `type IIA spacetime'), $\partial_y$ is along the
isometry direction of type IIB supergravity which is defined on
the spacetime with coordinates $x^\mu= (x^{\tilde{m}}, y)$ (which
may be called `type IIB spacetime'). Note that  nine of the ten
coordinates of `type IIA' and `type IIB' spacetimes are
identified, $\hat{x}^{\tilde{m}}=x^{\tilde{m}}$, ${\tilde{m}}=
0,1,\ldots , 8$.  Thus the intersection of type IIA and type IIB
spacetimes gives a nine--dimensional spacetime,
\begin{eqnarray}\label{9-0}
{\cal M}_{IIA}^{(10|0)} \cap {\cal M}_{IIB}^{(10|0)} = {\cal
M}^{(9|0)}  : \qquad x^{\tilde{m}}= (x^0, x^1 , \ldots , x^8) \; .
\end{eqnarray}
In other words, the Buscher rules (\ref{TrA})--(\ref{T-Bb}) imply
that (the bosonic fields) of type IIA and type IIB supergravity
are defined on different ten--dimensional surfaces in an
underlying 11 dimensional spacetime ({\it cf.} \cite{IBBJ})
\begin{eqnarray}\label{11-0}
{\cal M}^{(11|0)} : \quad X^M = (x^{\tilde{m}}, \hat{z}, y)
& \equiv & (\hat{x}^{\hat{\mu}} , y) \equiv (x^\mu , \hat{z}) \; , \\
\nonumber {\cal M}_{IIA}^{(10|0)} \subset {\cal M}^{(11|0)} :
\quad \hat{x}^{\hat{\mu}} = (x^{\tilde{m}}, \hat{z}) \; , \qquad
&&  {\cal M}_{IIB}^{(10|0)} \subset {\cal M}^{(11|0)} : \quad
{x}^\mu = (x^{\tilde{m}}, y) \; .
\end{eqnarray}

The field theoretical representation of T-duality was studied for
the  bosonic limit of supergravity in a number of papers (see,
refs. in \cite{IBBJ}). In particular, in \cite{simon} J. Simon
studied the T--duality transform of the Dp--brane actions
\cite{c0} in a purely {\it bosonic} supergravity background. In
this way he reproduced the above Buscher rules and also derived
the T-duality rules for the {\it RR (Ramond--Ramond)} gauge
potential of type IIA and type IIB supergravity,
\begin{eqnarray}\label{RR-RR}
\nonumber  RR\; \; type \; IIA & & RR\;\; type\; IIB
\\
 \hat{C}_{\hat{\mu}}(\hat{x})\, ,
 \hat{C}_{\hat{\mu}\hat{\nu}\hat{\rho}}(\hat{x})=
\hat{C}_{[\hat{\mu}\hat{\nu}\hat{\rho}]}(\hat{x})
 \,
& \longleftarrow^{T-duality}\longrightarrow & C_0(x)\, , \;
{C}_{{\mu}{\nu}}({x})= {C}_{[{\mu}{\nu}]}({x})\, ,
\\ \label{RR-RRD}
\begin{matrix}{ \hat{C}_{\hat{\mu}_1\ldots \hat{\mu}_7}(\hat{x})\, ,
 \hat{C}_{\hat{\mu}_1 \ldots \hat{\mu}_5}(\hat{x})\, , \cr
\hat{C}_{\hat{\mu}_1\ldots \hat{\mu}_9}(\hat{x})}\end{matrix}
 && \begin{matrix}{ C_{\hat{\mu}_1 \ldots \hat{\mu}_4}({x})= C_{[\hat{\mu}_1
\ldots \hat{\mu}_4]}({x})\cr
 C_{\hat{\mu}_1\ldots
\hat{\mu}_8}(x)\, , \, {C}_{\hat{\mu}_1\ldots \hat{\mu}_6}({x})\cr
 C_{\hat{\mu}_1\ldots
\hat{\mu}_{10}}(x)\,  }\end{matrix} \; ,
\end{eqnarray}
which had been  obtained several months before in \cite{MO98}.
Note that the {\it RR} T-duality rules from \cite{simon,MO98} involve
'dual' gauge fields (fields with dual field strengths) which are
higher rank antisymmetric tensors, as well as
$\hat{C}_{\hat{\mu}_1\ldots \hat{\mu}_9}(\hat{x})$ characteristic
for massive type IIA supergravity \cite{Romans} and the form
$C_{10}$ which is trivial when defined on the spacetime, but might
be nontrivial in superspace (see \cite{c0} and refs. therein)
\footnote{Note that the 4-form gauge field $C_{\hat{\mu}_1\ldots
\hat{\mu}_4}(x)$ of the type IIB supergravity has a self--dual
field strength, $F_5=dC_4+ \ldots = \ast F_5$. During almost 15
years this hampered the way to the covariant action of type IIB
theory. The problem was successfully resolved in \cite{IIBac} with
the use of the PST (Pasti--Sorokin--Tonin) approach \cite{PST}.}.

The Meessen--Ort\'{\i}n--Simon T-duality rules for the {\it RR} gauge fields have the
form
\begin{eqnarray}\label{Simon}
C^{(0)}&=& \hat{C}^{(1)}_{\hat{z}} \; ,
\\
C^{(2n)}_{y\tilde{m}_1\ldots \tilde{m}_{2n-1}} &=&
\hat{C}^{(2n-1)}_{\tilde{m}_1\ldots \tilde{m}_{2n-1}} +
{(2n-1)\over {\hat{g}_{\hat{z}\hat{z}}}}
\hat{C}^{(2n-1)}_{\hat{z}[\tilde{m}_1\ldots \tilde{m}_{2n-2}}
\hat{g}_{\tilde{m}_{2n-1}]\hat{z}} \; , \nonumber
\\
C^{(2n)}_{\tilde{m}_1\ldots \tilde{m}_{2n}}&=&
\hat{C}^{(2n+1)}_{\hat{z}\tilde{m}_1\ldots \tilde{m}_{2n}} + 2n
\hat{C}^{(2n-1)}_{[\tilde{m}_1\ldots \tilde{m}_{2n-1}}
\hat{B}_{\tilde{m}_{2n}]\hat{z}} + \nonumber \\
 &+& {2n(2n-1)\over {\hat{g}_{\hat{z}\hat{z}}}}
\hat{C}^{(2n-1)}_{\hat{z}[\tilde{m}_1\ldots \tilde{m}_{2n-2}}
\hat{B}_{\hat{z}\tilde{m}_{2n-1}}\hat{g}_{\tilde{m}_{2n}]\hat{z}}
\; , \nonumber  \\
\nonumber for  \; n& = & 1,2,3,4 \; , \; \; and
\\
C^{(10)}_{y\tilde{m}_1\ldots \tilde{m}_{9}} &=&
\hat{C}^{(9)}_{\tilde{m}_1\ldots \tilde{m}_{9}} + {9 \over
{\hat{g}_{\hat{z}\hat{z}}} }
\hat{C}^{(9)}_{\hat{z}[\tilde{m}_1\ldots \tilde{m}_{8}}
\hat{g}_{\tilde{m}_{9}]\hat{z}} \; . \nonumber
\end{eqnarray}

The T-duality rules for fermions
\begin{eqnarray}\nonumber
\nonumber  type \; IIA \qquad & & \qquad
type\; IIB \\ \nonumber
 \hat{\psi}_{\hat{\mu}}^{\alpha 1}(\hat{x}) \, ,
\hat{\psi}_{\hat{\mu}\alpha}{}^2(\hat{x}) \, ,
\; \hat{\lambda}_{\alpha}^1(\hat{x}) \, , \hat{\lambda}^{\alpha 2
}(\hat{x}) \, , &\leftarrow^{T-duality}\rightarrow& {\psi}_{{\mu}}^{\alpha 1}(x) \, ,
{\psi}_{{\mu}}^{\alpha 2}(x) \, , \;   {\lambda}_{\alpha}^1(x) \,
, {\lambda}_{\alpha}^2(x) \,
\end{eqnarray}
were recently obtained by Hassan \cite{hassan} by studying the map
between supersymmetry transformations of type IIA and type IIB
supergravity.

\section{Towards the superfield T-duality rules }

In superfield formulation of supergravity all the physical fields appear
as leading ($\hat{\theta } =0$ or $\theta =0$) components of some
superfields.
Moreover, one may say that differential forms
describing the physical fields, appear as the leading
($\hat{\theta } =0 \, , \, d\hat{\theta } =0$ or $\theta =0\, ,
d\theta=0$) terms in superforms of the superspace formulation of
$D=10$ type IIA and type IIB supergravity,
\begin{eqnarray}
\nonumber IIA \qquad & \qquad & \qquad IIB \qquad \\ \nonumber
\hat{e}^{\hat{a}} \equiv d\hat{x}^{\hat{\mu}}
\hat{e}_{\hat{\mu}}^{\hat{a}}(\hat{x}) =
\hat{E}^{\hat{a}}\vert_{\hat{\theta}=0=d\hat{\theta}} &\qquad &
{e}^{{a}} \equiv d{x}^{{\mu}} {e}_{\mu}^{{a}}({x}) =
\hat{E}^{\hat{a}}\vert_{{\theta}=0=d{\theta}}
\\ \nonumber
d\hat{x}^{\hat{\mu}} \hat{\psi}_{\hat{\mu}}^{\alpha 1}(\hat{x}) =
\hat{E}^{\alpha 1}\vert_{\hat{\theta}=0=d\hat{\theta}}\, ,  &
\quad & d{x}^{{\mu}} {\psi}_{{\mu}}^{\alpha 1}(x) =
\hat{E}^{\alpha 1}\vert_{{\theta}=0=d{\theta}}
 \\ \nonumber
d\hat{x}^{\hat{\mu}} \hat{\psi}_{\hat{\mu}\alpha}{}^2(\hat{x}) =
\hat{E}_{\alpha}{}^2 \vert_{\hat{\theta}=0=d\hat{\theta}}\, ,  &
\quad & d{x}^{{\mu}} {\psi}_{{\mu}}^{\alpha 2}(x) = {E}^{\alpha
2}\vert_{{\theta}=0=d{\theta}}
 \\ \nonumber
{d\hat{x}^{\hat{\mu}} \wedge \hat{x}^{\hat{\nu}} \over 2!}
\hat{B}_{\hat{\nu}\hat{\mu}}(\hat{x}) = \hat{B}_{2} (\hat{Z})
\vert_{...}\, , & \quad &
 {d{x}^{{\mu}} \wedge
{x}^{{\nu}}  \over (2)!} {B}_{{\nu}{\mu}}({x}) = {B}_{2}
({Z})\vert_{...} \; ,
 \\ \nonumber
 \hat{\phi}(\hat{x}) = \hat{\Phi}(\hat{Z}) \vert_{\hat{\theta}=0} &\qquad &
{\phi}({x}) = {\Phi}({Z})
\vert_{\theta=0} \\
\nonumber {d\hat{x}^{\hat{\mu}_1} \wedge \ldots \wedge
\hat{x}^{\hat{\mu}_{2n+1}}  \over (2n+1)!}
\hat{C}_{\hat{\mu}_{2n+1}\ldots  \hat{\mu}_{1}}(\hat{x}) = & \quad
&
 {d{x}^{{\mu}_1} \wedge \ldots \wedge
{x}^{{\mu}_{2n}}  \over (2n)!} {C}_{{\mu}_{2n}\ldots
{\mu}_{1}}({x}) =
\\
\nonumber = \hat{C}_{2n+1} (\hat{Z}) \vert_{\hat{\theta}=0=d\hat{\theta}}\, ,
& \quad & = {C}_{2n} ({Z})\vert_{{\theta}=0=d{\theta}} \: .
\end{eqnarray}
In this perspective the dilatons ($\phi(x)$ and $\hat{\phi}(\hat{x})$)
appear as
leading components of dilaton superfields (zero--form $\hat{\Phi}(\hat{Z})$
and $\Phi(Z)$) and
the spin $1/2$ fermions are
leading components of the covariant Grassmann derivatives of
these dilaton superfields
\begin{eqnarray}\label{1/2}
\nonumber \hat{\lambda}_{\alpha 1}(\hat{x}) \propto
\hat{\nabla}_{\alpha 1} \hat{\Phi}(\hat{Z})\vert_{\hat{\theta}=0}
& \qquad & {\lambda}_{\alpha 1}({x}) \propto {\nabla}_{\alpha 1}
{\Phi}({Z})  \vert_{{\theta}=0}\, , \nonumber \\
\hat{\lambda}^{\alpha}_2(\hat{x}) \propto \hat{\nabla}^{\alpha}_2
\hat{\Phi}(\hat{Z}) \vert_{\hat{\theta}=0} & \quad &
{\lambda}_{\alpha 2}({x}) \propto {\nabla}_{\alpha 2} {\Phi}({Z})
\vert_{{\theta}=0}\, ,
\end{eqnarray}

The problem 
was to find the complete set of T--duality rules relating the above superfield
supergravity potentials, or, equivalently, related superforms
\begin{eqnarray}\label{SUSY-td}
type \; IIA \; SUGRA  \qquad &    &
{} \quad type \; IIB \; SUGRA
\nonumber  \\  && \nonumber  \\
\left(\begin{matrix}
{{\hat{\Phi}}(\hat{Z}) \cr
\hat{E}^{\hat{a}}= d\hat{Z}^{\hat{M}} E_{\hat{M}}^{\hat{a}}(\hat{Z})  \cr
\hat{E}^{\alpha 1} =  d\hat{Z}^{\hat{M}} E_{\hat{M}}^{\alpha 1} (\hat{Z}) \cr
\hat{E}_{\alpha}{}^2 =
d\hat{Z}^{\hat{M}} E_{\hat{M}}{}_{\alpha}^2 (\hat{Z}) \cr
 \hat{B}_{2} (\hat{Z}) =
{d\hat{Z}^{\hat{M}} \wedge \hat{Z}^{\hat{N}} \over 2!}
\hat{B}_{\hat{N}\hat{M}}(\hat{Z})
\cr
\hat{C}(\hat{Z}) = \bigoplus^{4}_{n=0} \hat{C}_{2n+1} (\hat{Z})   \cr }
\end{matrix}\right) &
\begin{matrix}{\longleftrightarrow \cr
Superfield \cr T-duality \cr
\longleftrightarrow \cr  }
\end{matrix}  &
\left(\begin{matrix}
{{\Phi}({Z}) \cr {E}^{{a}}= d{Z}^{{M}} E_{{M}}^{{a}}({Z})  \cr
{E}^{\alpha 1} =  d{Z}^{{M}} E_{{M}}^{\alpha 1} ({Z})
\cr \hat{E}_{\alpha}{}^2  =  d{Z}^{{M}} E_{{M}}^{\alpha 2} ({Z})  \cr
 {B}_{2}({Z})={d{Z}^{{M}} \wedge {Z}^{{N}} \over 2!}
\hat{B}_{{N}{M}}({Z})  \cr
{C}({Z}) = \bigoplus^{5}_{n=0} {C}_{2n} ({Z})  \cr }
\end{matrix}\right) \; .
\end{eqnarray}
Where, following \cite{c0}, we collected all the RR superforms
of type IIA/IIB supergravity
 \begin{eqnarray}\label{rCIIA}
type~ IIA: \qquad & \hat{C}_{2n+1}
={1 \over (2n+1)!}
d\hat{Z}^{ {\hat{M}}_{2n+1}} \wedge \ldots \wedge
d\hat{Z}^{ {\hat{M}}_1}
\hat{C}^{(2n+1)}_{ {\hat{M}}_1\ldots
 {\hat{M}}_{2n+1}}(\hat{Z})\; ,
\\ \label{rCIIB}
type~ IIB: \qquad &
C_{2n}
={1 \over 2n!}
dZ^{{M}_{2n}} \wedge \ldots \wedge
dZ^{ {M}_1}C^{(2n)}_{ {M}_1\ldots  {M}_{2n}}(Z) \; .
\end{eqnarray}
in the formal sum of all odd/even differential forms on
superspace,
 \begin{eqnarray}
\label{rhC=hC+hC+}
type~ IIA: \qquad & \hat{C} =\hat{C}_1 \oplus \hat{C}_3  \oplus
\hat{C}_5 \oplus \hat{C}_7  \oplus \hat{C}_9 \equiv
 \bigoplus^{4}_{n=0} \hat{C}_{2n+1} (\hat{Z})  \; , & \qquad
\\
\label{rC=C+C+}
type~ IIB: \qquad & C=C_0 \oplus C_2 \oplus C_4  \oplus C_6  \oplus C_8
\oplus C_{10} \equiv  \bigoplus^{5}_{n=0} {C}_{2n} ({Z})
\;
& \qquad
\end{eqnarray}

In nineties there appeared a number of papers addressing this
problem (see \cite{siegel} and refs. in \cite{IBBJ}).  In
particular, in \cite{clps} the T-duality map of type IIA
superstring into the type IIB one and then back to type IIA
massive superstring (superstring in the background of Romans
massive type IIA supergravity \cite{Romans}) was studied  up to
quadratic order in the fermionic coordinate functions
$\tilde{\theta}(\tau,\sigma)$. In \cite{kulik} the  T--duality
rules for {\sl NS--NS superfields}
\begin{eqnarray}\label{NSNS-td}
type \; IIA \; SUGRA  \qquad &    &
{} \quad type \; IIB \; SUGRA
\nonumber  \\  && \nonumber  \\
\left(\begin{matrix}
{{\hat{\Phi}}(\hat{Z}) \cr
\hat{E}^{\hat{a}}= d\hat{Z}^{\hat{M}} E_{\hat{M}}^{\hat{a}}(\hat{Z})  \cr
 \hat{B}_{2} (\hat{Z}) =
{d\hat{Z}^{\hat{M}} \wedge \hat{Z}^{\hat{N}} \over 2!}
\hat{B}_{\hat{N}\hat{M}}(\hat{Z})
\cr  }
\end{matrix}\right) &
\begin{matrix}{ {}^{T-duality} \cr
\longleftrightarrow \cr  }
\end{matrix}  &
\left(\begin{matrix}
{\Phi(Z) \cr  {E}^{{a}}= d{Z}^{{M}} E_{{M}}^{{a}}({Z})  \cr
 {B}_{2} ({Z})= {d{Z}^{{M}} \wedge {Z}^{{N}} \over 2!}
\hat{B}_{{N}{M}}({Z})  \cr }
\end{matrix}\right) \; .
\end{eqnarray}
and fermionic supervielbeins
\begin{eqnarray}\label{FF-td}
\left(\begin{matrix}
{ \hat{E}^{\alpha 1} =  d\hat{Z}^{\hat{M}} E_{\hat{M}}^{\alpha 1} (\hat{Z})
\cr
\hat{E}_{\alpha}{}^2 =
d\hat{Z}^{\hat{M}} E_{\hat{M}}{}_{\alpha}^2 (\hat{Z}) \cr }
\end{matrix}\right)
 &
\longleftarrow
{}^{T-duality}
\longrightarrow &
\left(\begin{matrix}
{{E}^{\alpha 1} =  d{Z}^{{M}} E_{{M}}^{\alpha 1} ({Z})
\cr \hat{E}_{\alpha}{}^2  =  d{Z}^{{M}} E_{{M}}^{\alpha 2} ({Z})  \cr }
\end{matrix}\right) \; .
\end{eqnarray}
were found by studying the relation between complete
type IIA and type IIB superstring actions and their $\kappa$--symmetries.
However, this approach did not allow to find the T--duality rules for
Ramond--Ramond (RR) {\sl superfield potentials}
\begin{eqnarray}\label{RR-td}
\hat{C}(\hat{Z}) = \bigoplus^{4}_{n=0} \hat{C}_{2n+1} (\hat{Z})
 &
\longleftarrow
{}^{T-duality}
\longrightarrow &
C(Z) = \bigoplus^{5}_{n=0} C_{2n} (Z)
\end{eqnarray}
and required significant efforts to extract the transformation rules for the
{\sl components} of the RR field strengths from Bianchi identities.

\section{
Derivation of  superfield T-duality rules
(\ref{SUSY-td}).}

The complete set of superfield T--duality rules were obtained in
\cite{IBBJ}. They follow from the relation between
the complete $\kappa$--symmetric actions
for Dirichlet superbranes in type IIA and type IIB
supergravity backgrounds
and subsequent study of the
exchange between the type IIA and IIB superspace supergravity constraints.
Namely, in the first stage, the comparison of the
type IIA super--D$(p+1)$--brane and type IIB
super--Dp--brane actions \cite{c0}
which are known to be related by T--duality \cite{Polch0,Polch},
provides  the T--duality transformation rules for the {\it bosonic} superforms
of type IIB resp. IIA supergravity
(including all RR superforms), Eqs. (\ref{NSNS-td}), (\ref{RR-td}).
Then, in a second stage,
substituting these rules into the
superspace torsion constraints and the constraints
on NS--NS field strengths of type IIA and type IIB supergravities
\cite{HW84,CGO87,c0}, one can derive the T--duality rules for the remaining
(fermionic) supervielbein forms.

It turns out that the
T-duality transformation rules for the bosonic superforms, which
can be obtained from the comparison of
the super--Dp--brane actions ({\it i.e.} by the
superfield generalization of the {\sl method} of Ref. \cite{simon}),
can be reproduced as well by a straightforward
superfield (superform) generalization of the
{\sl final results} of
Ref. \cite{simon}.
\footnote{
Such a simple possibility to reproduce the
superfield results form the component ones can be regarded as a reflection
of the existence of the `rheonomic'
(group manifold) approach to supergravity \cite{rheo}
which allows to lift the component equations
(written
in terms of differential forms on spacetime) to the superspace equations
for superforms. This is also natural in a view of
recent observation \cite{BdAIL} that superfield description of the
dynamical supergravity--superbrane interacting system (still hypothetical
for $D=10,11$) is gauge equivalent to a more simple dynamical system
described by the sum of the standard (component) supergravity action
and the action for pure bosonic brane
(the pure bosonic limit of the original superbrane action).}

In \cite{IBBJ} we used such a shortcut.
By substituting the NS--NS T--duality rules thus obtained
into the superspace torsion constraints and into the constraints
on NS--NS field strengths
\cite{HW84,CGO87,c0}, we have derived the T--duality rules for fermionic
supervielbein forms (in Einstein frame).
Finally, we described in \cite{IBBJ} the verification
of the consistency of the complete set of
T--duality rules thus obtained with the superspace constraints for RR
superform field strengths \cite{c0}.

\subsection{
T--duality as an operation acting on differential forms and
underlying superspace ${\cal M}^{(11|32)}$.}

Our approach \cite{IBBJ} treats
{\it T--duality as
an operation which acts on differential
forms in superspace rather than on the superspace coordinates}.
\footnote{Such a possibility is guaranteed by (super)diffeomorphism invariance
of (superspace super)gravity, {\it i.e.} by
its  gauge symmetry under arbitrary
changes of local coordinate system (in superspace).
(Super)diffeomorphism invariance
 allows one to replace any coordinate transformations by the equivalent
transformations  of the supergravity (super)fields (see, {\it e.g.},
\cite{BdAI} and refs. therein).  However, such `picture changing' allows
to overcome the problems that blocked the way to superfield
T--duality rules (see e.g. \cite{simon2}).}
This treatment of T-duality
allows one to identify all but one bosonic and {\sl all} the fermionic
coordinates of {\sl curved}
type IIA and type IIB superspace
\footnote{The possibility of identification of the fermionic coordinates
of {\sl curved} type IIA and type IIB superspaces,
$\hat{\theta}^{{\hat{\mu}}}={\theta}^{{{\mu}}}$,
 Eqs. (\ref{MIIA}), (\ref{MIIB}), would not seem surprising if
one remembers that the fermionic coordinates of a general
{\sl curved} superspace do not carry any chirality.
Their indices $\hat{{\mu}}$ and ${\mu}$
are {\sl not} the spinor indices of the Lorentz group;
$\hat{\theta}^{{\hat{\mu}}}$ and ${\theta}^{{{\mu}}}$ are
rather transformed by the general superdiffeomorphism symmetry.
The {\sl chirality}, which is used  to distinguish the
type IIA and type IIB
is a characteristic
of the fermionic supervielbein 1--forms
($\hat{E}^{\alpha 1},
\hat{E}_{\alpha}^2$ and  $\hat{E}^{\alpha 1}, \hat{E}^{\alpha 2}$)
which do carry $SO(1,9)$ spinor indices.
Only in the flat superspace limits, when one takes the fermionic
supervielbein to be  derivatives of the fermionic coordinates,
the chiral structure, together
with the definite spinor representation of the Lorentz group,
applies to the fermionic coordinates of flat superspace.}, i.e.
  \begin{eqnarray}
\label{MIIA}
 type~IIA: & \qquad     {\cal M}_{IIA}^{(10|32)} \; : & \quad
    \hat{Z}^{ {\hat{M}}} =
 (\tilde{{Z}}{}^{ {\tilde{M}}}, \hat{z})\; ,
\\ \label{MIIB}
  type~IIB: & \qquad   {\cal M}_{IIB}^{(10|32)} \; :  & \quad
{Z}^{ {{M}}}=
 (\tilde{Z}^{\tilde{M}}, y )\; .
\end{eqnarray}
In other words, the intersection of curved type IIA and
type IIB superspaces, ${\cal M}_{IIA}^{(10|32)}$ and
${\cal M}_{IIB}^{(10|32)}$, defines some $D=9$, $N=2$ superspace
${\cal M}^{(9|32)}$ ({\it cf.} (\ref{9-0}))
\begin{eqnarray}
\label{M9}
 {\cal M}_{IIA}^{(10|32)} \cap  {\cal M}_{IIB}^{(10|32)} &=&
 {\cal M}^{(9|32)} \;  \\
\label{ZM9}
{\cal M}^{(9|32)} \; :
\tilde{Z}^{ \tilde{M}}
      &\equiv& \left(
          \tilde{X}^{\tilde{m}},
       {\theta}^{\mu}\right) \; ,
\quad
 {\tilde{m}}= 0, \ldots , 8 \; , \qquad
 {{\mu}}= 1, \ldots , 32 \; . \qquad
\end{eqnarray}
 Moreover, this point of view
makes transparent that type IIA and type IIB  theories
with isometries $\partial_{\hat{z}}$ and $\partial_{y}$
can be defined
on the hypersurfaces $\hat{z}=0$ and $y=0$ of an {\it underlying superspace}
${\cal M}^{(11|32)}$ with $11$ bosonic and $32$ fermionic coordinates,
\begin{eqnarray}
\label{M11}
  & {\cal M}^{(11|32)} \; : \quad
   (\tilde{{Z}}{}^{ {\tilde{M}}}, \; y, \; \hat{z}\;) \; .
\end{eqnarray}

\section{
Superfield T--duality rules}

The superfield T--duality rules are simplest when written
in terms of the supervielbein forms adapted to the
isometry, i.e. obeying the
superfield generalization  of the Kaluza--Klein type ansatz familiar from
dimensional reductions of supergravity theories
\cite{ansatz}:
\begin{eqnarray}\label{KKA}
\hbox{type IIA}\,: \hat{E}^{\hat{a}}&=& (\hat{E}^{\tilde{a}},
\hat{E}^{\#}) \,, \qquad \hat{E}^{ {\tilde{a}}}= \hat{E}^{
  {\tilde{a}}(-)}= d\tilde{Z}^{\tilde{M}} \hat{E}^{ {\tilde{a}}}_{\tilde{M}}
(\tilde{Z}) \; , \qquad
\\ \nonumber {\hat{a}}=0, \ldots 9, \quad && a= 0, \ldots 9, \qquad
\tilde{a} =0, \ldots , 8
\\
\label{KKB}
\hbox{type IIB} \,: E^a &=& ({E}^{\tilde{a}}, {E}^{*}) \,, \qquad
     {E}^{ {\tilde{a}}}= {E}^{ {\tilde{a}}(-)} = d\tilde{Z}^{\tilde{M}}
{E}^{{\tilde{a}}}_{\tilde{M}}(\tilde{Z}) \,, \qquad
\end{eqnarray}
i.e. $\hat{E}_{\hat{z}}^{\tilde{a}}=0$, ${E}_{y}^{\tilde{a}}=0$,
and all the nonvanishing component-superfields  depending only
on the coordinates
$\tilde{Z}^{\tilde{M}}$ of
nine-dimensional superspace ${\cal M}^{(9|32)}$, Eq. (\ref{M9}).

T-duality rules for NS-NS superfields read \cite{IBBJ}
\begin{equation}\label{Ts1}
e^{{\Phi(\tilde{Z})\over 4}} E_{\tilde{M}}^{ {\tilde{a}}} =
e^{{\hat{\Phi}(\tilde{Z})\over 4}}\hat{E}_{\tilde{M}}^{ {\tilde{a}}} \,, \qquad
e^{{\Phi\over 4}} E_y^{ {*}}= {1\over e^{{\hat{\Phi}\over 4}}
  \hat{E}_{\hat{z}}^{ {\#}}}\,, \qquad
e^{{\Phi\over 4}} E_{\tilde{M}}^{{*}}=
{\hat{B}_{\hat{z}\tilde{M}}  \over e^{{\hat{\Phi}\over 4}}
  \hat{E}_{\hat{z}}^{ {\#}}} \,,
\end{equation}
\begin{equation}\label{Ts1P}
e^{{\Phi}(\tilde{Z})} = {e^{\hat{\Phi}(\tilde{Z})}\over
  e^{{\hat{\Phi}\over 4}} \hat{E}_{\hat{z}}^{ {\#}}} \,, \qquad
\end{equation}
\begin{equation}\label{Ts2}
B_{y\tilde{M}} = {\hat{E}^{ {\#}}_{\tilde{M}}\over
\hat{E}_{\hat{z}}^{ {\#}}} \,,
\qquad
B_{\tilde{M}\tilde{N}}= \hat{B}_{\tilde{M}\tilde{N}}
- {2 \over \hat{E}_{\hat{z}}^{ {\#}}}  \hat{B}_{\hat{z}[\tilde{M}}
 \hat{E}^{ {\#}}_{\tilde{N}\} }\; .
\end{equation}

The T--duality rules for fermionic supervielbeins are
\begin{eqnarray}\label{Tsf19}
e^{-{1\over 8}{\Phi}} {{E}_y^{{\beta}1} \over E_y^*} &=& - \,
e^{-{1\over 8}\hat{\Phi}}\; \left( {\hat{E}_{\hat{z}}^{ {{\beta}1}}\over
  \hat{E}_{\hat{z}}^{ {\#}}} + {i \over 4} \tilde{\sigma}^{ {\#}}{}^{
  {\beta} {\gamma}} \hat{\nabla}_{ {\gamma}1}\hat{\Phi}- {i \over 8}
\tilde{\sigma}^{ {\#}}{}^{ {\beta} {\gamma}} \hat{\nabla}_{ {\gamma}1}
\ln \left(e^{\hat{\Phi}\over 4} E_{\hat{z}}^{\#} \right)\right),
\qquad \\
\label{Tsf29}
e^{-{1\over 8}{\Phi}} {{E}_y^{ {\beta}2}\over E_y^*} &=& \; e^{-
  {1\over 8}\hat{\Phi}}\; \tilde{\sigma}^{ {\#} {\beta} {\gamma}} \;
\left({\hat{E}_{\hat{z}}{}_{\gamma}^2 \over \hat{E}_{\hat{z}}^{ {\#}}}
+ {i \over 4} {\sigma}^{\#}_{ {\beta} {\gamma}} \hat{\nabla}^{
  {\gamma}}_2\hat{\Phi} - {i \over 8} {\sigma}^{\#}_{ {\beta}
  {\gamma}} \hat{\nabla}^{ {\gamma}}_2 \ln \left({e^{\hat{\Phi}\over
    4}} E_{\hat{z}}^{\#}\right) \right), \qquad \\
\label{Tsf1-}
e^{{1\over 8}{\Phi}} {E}^{ {\beta}1[-]} &=& \; e^{{1\over
    8}\hat{\Phi}}\; \left(\hat{E}^{{\beta}1[-]} - {i \over 8}
\hat{E}^{ {\tilde{a}}(-)} \tilde{\sigma}_{ {\tilde{a}}}{}^{ {\beta}
  {\gamma}} \hat{\nabla}_{ {\gamma}1} \ln \left({e^{\hat{\Phi}\over
    4}} E_{\hat{z}}^{\#} \right) \right), \qquad \\
\label{Tsf2-}
e^{{1\over 8}{\Phi}} {E}^{ {\beta}2[-]} &=& \; e^{{1\over
    8}\hat{\Phi}}\; \tilde{\sigma}^{ {\#} {\beta} {\gamma}} \; \left(
\hat{E}_{ {\gamma}}^{2[-]} - {i \over 8} \hat{E}^{{\tilde{a}}(-)}
    {\sigma}_{\tilde{ {a}} {\beta} {\gamma}} \hat{\nabla}^{
      {\gamma}}_2 \ln \left({e^{\hat{\Phi}\over 4}}
    E_{\hat{z}}^{\#}\right)\right)\; ,  \qquad
\end{eqnarray}
where
 \begin{eqnarray}\label{EfIIA-}
&&
\hat{E}^{ {{\alpha}1}[-]}=
d\tilde{Z}^{\tilde{M}} \left(
{E}_{\tilde{M}}^{{\alpha}1} -
     \hat{E}_{\tilde{M}}^{\#} {\hat{E}_{\hat{z}}^{{\alpha}1}\over
\hat{E}_{\hat{z}}^{\#}} \right)\; ,
\qquad
\hat{E}^{2[-]}_{ {{\alpha}}} = d\tilde{Z}^{\tilde{M}} \left(
\hat{E}_{\tilde{M}}{}^{2}_{ {{\alpha}}} -
     \hat{E}^{ {\#}}_{\tilde{M}} {\hat{E}_{\hat{z}}{}_{\alpha}^2\over
\hat{E}_{\hat{z}}^{\#}}\right)\; , \qquad
\\ \label{EfIIB-}
&&
{E}^{{\alpha}1[-]}= d\tilde{Z}^{\tilde{M}} \left(
                  E^{ {{\alpha}}1}_{\tilde{M}} -
     {E}^{ {*}}_{\tilde{M}} {{E}_y^{ {{\alpha}}1}\over E_y^*} \right) \; ,
\qquad
{E}^{{\alpha}2[-]}=  d\tilde{Z}^{\tilde{M}} \left(
                  E_{\tilde{M}}^{ {{\alpha}}2} -
     {E}^{ {*}}_{\tilde{M}} {{E}_y^{ {{\alpha}}2}\over E_y^*}\right)  \; .
\end{eqnarray}
One may rewrite Eqs. (\ref{Tsf1-}),  (\ref{Tsf2-}) in terms of 
${E}^{{\alpha}1(-)}= d\tilde{Z}^{\tilde{M}} 
                  E^{ {{\alpha}}1}_{\tilde{M}}$, 
${E}^{{\alpha}2(-)}=  d\tilde{Z}^{\tilde{M}} 
                  E_{\tilde{M}}^{ {{\alpha}}2}$ 
and their type IIA counterparts.
However, the notion of $^{[-]}$ components of differential forms, 
which implies separation of the terms proportional to 
$E^*$ and $\hat{E}^{\#}$ (see Eqs. (\ref{KKA}), (\ref{KKB})), 
rather than the ones proportional to $dy$ and $d\hat{z}$,  is quite suggestive.
In particular, with this notation, the fermionic T--duality transformation 
rules  
(\ref{Tsf19})-- (\ref{Tsf2-}) do not involve any contribution from 
the NS--NS superfields 
enclosed in $B_{2}$ and $\hat{B}_2$. Such contributions appear when 
one rewrites (\ref{Tsf1-}),  (\ref{Tsf2-})  in terms of 
${E}^{{\alpha}1(-)}$ etc. (see \cite{IBBJ}), but, taking in mind 
the original form of Eqs. (\ref{Tsf1-}),  (\ref{Tsf2-}), one immediately 
clarifies their origin in the T--duality rules 
for bosonic superfields ${E}^{ {*}}_{\tilde{M}}(\tilde{Z})$ and 
$\hat{E}^{{\#}}_{\tilde{M}}(\tilde{Z})$, Eqs. (\ref{Ts1}), (\ref{Ts2}).

The above set of T--duality rules from \cite{IBBJ}
coincides with the ones from \cite{kulik} after passing to the
so--called String frame
and to the supervielbeins which are not adapted to isometries
(see \cite{IBBJ} for a detailed comparison).

However, our approach allowed us to derive as well the
T-duality rules for the RR superform potentials \cite{IBBJ}. They are
\begin{eqnarray}\label{Ts3}
& C^{(0)} =  \hat{C}^{(1)}_{\hat{z}}\,,
\nonumber \\
& C^{(2n)}_{y\tilde{M}_1\ldots \tilde{M}_{2n-1}}  =
- \hat{C}^{(2n-1)}_{\tilde{M}_1\ldots \tilde{M}_{2n-1}} +
{2n-1\over \hat{E}_{\hat{z}}^{ {\#}}}\,
\hat{E}^{ {\#}}_{[\tilde{M}_1|}
\hat{C}^{(2n-1)}_{\hat{z}|\tilde{M}_2\ldots \tilde{M}_{2n-1}\} }
\,, \qquad
\\
& C^{(2n)}_{\tilde{M}_1\ldots \tilde{M}_{2n}}
=
\hat{C}^{(2n+1)}_{\hat{z}\tilde{M}_1\ldots \tilde{M}_{2n}} +
2n  \hat{B}_{\hat{z}[\tilde{M}_1}
\hat{C}^{(2n-1)}_{\tilde{M}_2\ldots \tilde{M}_{2n}\} }
+ {2n (2n-1)\over   \hat{E}_{\hat{z}}^{ {\#}}}
\hat{B}_{\hat{z}[\tilde{M}_1} \hat{E}^{ {\#}}_{\tilde{M}_2|}
\hat{C}^{(2n-1)}_{\hat{z}| \tilde{M}_3\ldots \tilde{M}_{2n}\} }\; .
\nonumber
\end{eqnarray}

\subsection{
Superfield T--duality rules in differential form notation.}

The above  Eqs. (\ref{Ts3})
may be collected in compact expressions for the formal sums of all
type IIA and all type IIB superforms, (\ref{rhC=hC+hC+}) and (\ref{rC=C+C+}),
\begin{equation}
\label{rC=-r1}
C= i_{\hat{z}}\hat{C} + (dy +i_{\hat{z}} \hat{B}_2) \wedge
\left(\hat{C}^{(-)} - \, {\hat{E}^{ {\#}(-)}\over \hat{E}_{\hat{z}}^{
    {\#}}}\, \wedge i_{\hat{z}} \hat{C} \right),
\end{equation}
or
\begin{equation}\label{rhC=r1}
\hat{C} = - i_y C + (d\hat{z}+ i_yB_2)\wedge \left(C^{(-)}- i_y C
\wedge \, {{E}^{ {*}(-)}\over {E}_y^{ {*}}}\, \right)\; .
\end{equation}
Here the contractions $i_{\hat{z}}$
and the `minus' components $\hat{C}^{(-)}$
are defined by
\begin{eqnarray}\label{IIA-iOmd}
& i_{\hat{z}}\hat{\Omega}_q := {1\over (q-1)!} d\tilde{Z}^{\tilde{M}_{q-1}}
\wedge \ldots \wedge
d\tilde{Z}^{\tilde{M}_1}
\hat{\Omega}_{\hat{z} \tilde{M}_1\ldots \tilde{M}_{q-1}}(\tilde{Z})
\; , \\
\label{IIA-Omd-}
& \hat{\Omega}^{(-)}_q := {1\over q!} d\tilde{Z}^{\tilde{M}_{q}}
\wedge \ldots \wedge
d\tilde{Z}^{\tilde{M}_1}
\hat{\Omega}_{\tilde{M}_1\ldots \tilde{M}_{q}}(\tilde{Z})
\;
\end{eqnarray}
for any q--forms
$\hat{\Omega}_q = {1\over q!} d\hat{Z}^{M_q} \wedge \ldots \wedge
d\hat{Z}^{M_1}
\hat{\Omega}_{M_1\ldots M_q}(\hat{Z})$
in type IIA superspace; in the same way one defines
$i_y$ and $(-)$ components in the type IIB case.

The similar expression may be derived for the T--duality rules
for NS--NS superforms
\begin{equation}
\label{TBsg-r1}
{{B}}_2= \hat{{B}}_2 - (dy + i_{\hat{z}} \hat{B}_2) \wedge
\left(d\hat{z}+ {\hat{E}^{ {\#}(-)} \over \hat{E}_{\hat{z}}^{
    {\#}}}\right) + dy \wedge d\hat{z}\,,
\end{equation}
or
\begin{equation}
\label{rTdB2-r2}
B_2 = \hat{B}_2 - \, {1\over {E}_y^{ {*}}} \, E^{ {*}} \wedge
\hat{E}^{ {\# }} \, {1\over \hat{E}_{\hat{z}}^{ {\#}}}\, + dy \wedge
d\hat{z} \, ,
\end{equation}
and
for fermionic supervielbeins:
\begin{eqnarray}
\label{Td-EhE1r!}
e^{{1\over 8}{\Phi}} ({E}^{ {\beta}1} - {i \over 8} {E}^{ {a}}
\tilde{\sigma}_{ {a}}{}^{ {\beta} {\gamma}} {\nabla}_{
  {\gamma}1}{\Phi}) &=& e^{{1\over 8}\hat{\Phi}}\; (\hat{E}^{
  {\beta}1} - {i \over 8} \hat{E}^{ {a}} \tilde{\sigma}_{ {a}}{}^{
  {\beta} {\gamma}} \hat{\nabla}_{ {\gamma}1}\hat{\Phi}) -
\\  && \hspace{-2cm}
-\, e^{{1\over 8}\hat{\Phi}}\; \left(\hat{E}^{ {\#}} + e^{{1\over
    4}(\Phi -\hat{\Phi})} {E}^{ {*}}\right) \; \left( {
  \hat{E}_{\hat{z}}^{ {\beta}1}\over \hat{E}_{\hat{z}}^{ {\#}}} + {i
  \over 8} \tilde{\sigma}^{ {\#}}{}^{ {\beta} {\gamma}} \hat{\nabla}_{
  {\gamma}1}\hat{\Phi}\right),
\nonumber\\
\label{Td-EhE2r!}
e^{{1\over 8}{\Phi}} ({E}^{ {\beta}2} - {i \over 8} {E}^{ {a}}
\tilde{\sigma}_{ {a}}{}^{ {\beta} {\gamma}} {\nabla}_{
  {\gamma}2}{\Phi}) &=& e^{{1\over 8}\hat{\Phi}}\; \tilde{\sigma}^{
  {\#} {\beta} {\gamma}} \; ( \hat{E}_{ {\gamma}}^2 - {i \over 8}
\hat{E}^{ {a}} {\sigma}_{{ {a}} {\beta} {\gamma}} \hat{\nabla}^{
  {\gamma}}_2\hat{\Phi}) -
\\ && \hspace{-2cm}
-\, e^{{1\over 8}\hat{\Phi}}\, \left(\hat{E}^{ {\#}} - e^{{1\over
    4}(\Phi -\hat{\Phi})} {E}^{ {*}}\right) \, \tilde{\sigma}^{ {\#}
  {\beta} {\gamma}} \left({\hat{E}_{\hat{z}}{}_{ {\gamma}}^2 \over
  \hat{E}_{\hat{z}}^{ {\#}}} + {i \over 8} {\sigma}^{\#}_{ {\beta}
  {\gamma}} \hat{\nabla}^{ {\gamma}}_2\hat{\Phi}\right)\; .
\nonumber
\end{eqnarray}
Just this compact form of the fermionic T--duality rules allowed us
to check \cite{IBBJ} the consistency of the full set of the T--duality rules
with the complete set of superfield supergravity constraints
\cite{HW84,CGO87,c0},
including the ones for the RR superfield strengths.

\section{
A brief conclusion}

More details and a discussion can be found in the original paper  \cite{IBBJ}.
Our approach  can be also extended to
the more complicated  $SO(n,n)$ T--duality provided
the superfield generalization
of the Kaluza--Klein  type ansatz for the dimensional reduction
down to $d=10-n$ dimensions is elaborated for this case.
The results of our study
clarify the relationship of T--duality with superfield
formulations of supergravity and, as  we hope, might provide new
insights in M--theory.

\end{document}